\documentclass[twocolumn]{aastex6}
\usepackage{amsmath}

\newcommand{\fesc}{f_{{\rm esc}}}
\newcommand{\tauel}{\tau_{{\rm el}}}
\newcommand{\qhii}{Q_{{\rm HII}}}
\newcommand{\muv}{M_{{\rm UV}}}
\newcommand{\muvlim}{M_{{\rm UV,lim}}}
\def \oii{[O{\tiny\,II}]}
\def \oiii{[O{\tiny\,III}]}
\def \nii{[N{\tiny\,II}]}
\def \hii{H{\tiny\,II}}
\def \hbeta{H$\beta$}
\def \halpha{H$\alpha$}
\def \Msol{{\rm M}_{\odot}}

\def \logm{\log(M/\Msol)}
\def \lya{Ly$\alpha$~}
\def \ewha{{\rm EW}({\rm H}\alpha)}
\def \xiion{ \xi_{{\rm ion}} }

\begin{document}


\title{Revisiting the Lyman Continuum escape fraction crisis: Predictions for $z>6$ from local galaxies}

%% Use \author, \affil, plus the \and command to format author and affiliation 
%% information.  If done correctly the peer review system will be able to
%% automatically put the author and affiliation information from the manuscript
%% and save the corresponding author the trouble of entering it by hand.
%%
%% The \affil should be used to document primary affiliations and the
%% \altaffil should be used for secondary affiliations, titles, or email.

%% Authors with the same affiliation can be grouped in a single
%\author{Andreas L. Faisst\altaffilmark{1,2}}
\author{Andreas L. Faisst}

%\affil{American Astronomical Society \\
%2000 Florida Ave., NW, Suite 300 \\
%Washington, DC 20009-1231, USA}

%% Use the \and command so offset the last author.
%\and

%\author{Jeff Lewandowski\altaffilmark{5}}
%\affil{IOP Publishing, Washington, DC 20005}


\affil{Infrared Processing and Analysis Center, California Institute of Technology, Pasadena, CA 91125, USA}
%\affil{Cahill Center for Astronomy and Astrophysics, California Institute of Technology, Pasadena, CA 91125, USA}

%\altaffiltext{1}{AAS Journals Data Scientist}
%\altaffiltext{2}{greg.schwarz@aas.org}
%\altaffiltext{3}{AAS Journals Associate Editor-in-Chief}
%\altaffiltext{4}{AAS Director of Publishing}
%\altaffiltext{5}{IOP Senior Publisher for the AAS Journals}

\email{afaisst@ipac.caltech.edu; Twitter: @astrofaisst}
%\altaffiltext{}{Twitter: @astrofaisst}

\begin{abstract}

	The intrinsic escape fraction of ionizing Lyman continuum photons ($\fesc$) is crucial to understand whether galaxies are capable of reionizing the neutral hydrogen in the early universe at $z>6$. Unfortunately, it is not possible to access $\fesc$ at $z>4$ with direct observations and the handful of measurements from low redshift galaxies consistently find $\fesc<10\%$, while at least $\fesc \sim 10\%$ is necessary for galaxies to dominate reionization.
	Here, we present the first empirical prediction of $\fesc$ at $z>6$ by combining the (sparsely populated) relation between \oiii/\oii~and $\fesc$ with the redshift evolution of \oiii/\oii~as predicted from local high-z analogs selected by their \halpha~equivalent-width.
	We find $\fesc=5.7_{-3.3}^{+8.3}\,\%$ at $z=6$ and $\fesc=10.4_{-6.3}^{+15.5}\,\%$ at $z=9$ for galaxies with $\logm \sim 9.0$ (errors given as $1\sigma$). However, there is a negative correlation with stellar mass and we find up to $50\%$ larger $\fesc$ per $0.5\,{\rm dex}$ decrease in stellar mass. The population averaged escape fraction increases according to $\fesc = f_{{\rm esc,0}}\left((1+z)/3\right)^\alpha$, with $ f_{{\rm esc,0}}=2.3\pm0.05$ and $\alpha=1.17\pm0.02$ at $z>2$ for $\logm \sim 9.0$. 
	With our empirical prediction of $\fesc$  (thus fixing an important previously unknown variable) and further reasonable assumption on clumping factor and the production efficiency of Lyman continuum photons, we conclude that the average population of galaxies is just capable of reionizing the universe by $z\sim6$.

\end{abstract}




\section{Introduction} \label{sec:intro}

	A major phase transition in the early universe takes place during the \textit{Epoch of Reionization} (EoR), in which hydrogen in the inter-galactic medium (IGM) is transformed from a neutral to an ionized state. The EoR is closely connected to the formation of the first galaxies and thus the study of its evolution in time and space is important to understand galaxy formation in the early universe.
	
	The study of absorption due to intervening neutral hydrogen in the IGM in ultra-violet (UV) spectra of quasars allow us to pinpoint the end of the EoR (i.e., the time when the universe is fully ionized) to $z\sim6$ \citep[][]{FAN06,MCGREER11,MORTLOCK11}.
	Furthermore, the rapid decrease in the fraction of star-forming high redshift galaxies with \lya emission at $z>6$ suggest that the universe got ionized very quickly on timescales of only a couple $100\,{\rm Myrs}$ between $z\sim6-10$ \citep[e.g.,][]{STARK10,ONO12,SCHENKER13,MATTHEE14,FAISST14,ROBERTSON15}.
	In addition to these direct observations, the temperature fluctuations in the cosmic microwave background (CMB) allow the measurement of the integrated density of free electrons from $z=0$, through the EoR, to $z\sim1100$ when the CMB emerged. Recent measurements suggest $\tau_{e}=0.055\pm 0.009$ and constrain the end of the EoR to $7.8 \lesssim z_{\rm ion} \lesssim 8.8$ assuming an immediate ionization of hydrogen \citep{PLANCK16}.
	
	Although such observations are able to reveal the time frame of the EoR, we are mostly tripping in the dark about the origin of the dominant ionizing sources. Quasars and star-forming galaxies are currently the competing players for providing energetic photons to ionize hydrogen at $z>6$.
	However, because of the suggested sharp decline in the number density of quasars with increasing redshift at $z>6$, they likely do not dominate the budget of radiation needed to ionize hydrogen\footnote{However, they contribute to the reionization of Helium at $z\sim3$ \citep[see also][]{MADAU15}.}\citep[e.g.,][]{MASTERS12,PALANQUEDELABROUILLE13}.
	On the other hand, the overall number density of UV emitting, faint star-forming galaxies has only slightly dropped between $6 < z < 9$ \citep[][]{TACCHELLA13,SCHENKER13,OESCH14,BOUWENS15b,MASON15}. Furthermore, studies of faint, lensed galaxies show the continuation of the UV luminosity function (LF) to very faint magnitudes \citep[][]{ALAVI14,LIVERMORE16}, thus providing an important number of galaxies needed for reionization. 
	
	The redshift evolution of the volume fraction of ionized hydrogen ($\qhii$) and the integral of the electron scattering optical depth ($\tauel(z)$, the integrated density of free electrons to redshift $z$) allow us to test whether galaxies are actually capable of reionizing the universe \citep[e.g.,][]{FINKELSTEIN12a,KUHLEN12,BOUWENS15a,ROBERTSON15,PRICE16}.
	Unfortunately, the determination of $\qhii(z)$ and $\tauel(z)$ involves several properties of galaxies and their environment, which cannot be measured directly or have to be accessed via cosmological simulations.
	In detail, these dependencies are the faint end slope of the UV LF and its cut-off magnitude ($\muvlim$), the clumping of hydrogen in the IGM ($C$), the Lyman continuum (LyC) photon production efficiency ($\xiion$), and the intrinsic escape fraction of ionizing LyC photons ($\fesc$).
	We have a good handle on $\muvlim$ from lensing (see above), and good estimates on $\xiion$ at $z\sim5$ \citep[e.g.,][]{BOUWENS15}\footnote{Note that this measurement depends on the assumed stellar populations. Specifically, the inclusion of binary stellar populations may lead to significantly higher $\xiion$ \citep[][]{MA16,STANWAY16,STEIDEL16,WILKINS16}. } , and can provide a reasonable range in $C$ from cosmological simulations \citep[e.g.,][]{FINLATOR12}. In contrast, $\fesc$ is puzzling and unfortunately directly affecting $\qhii$ and $\tauel$ and therefore our picture of galaxies during reionization.
	
	With only $\fesc$ as free parameter, different studies suggest that $\fesc=10\%-20\%$ at $z>6$ is necessary for galaxies to fully ionize the universe \citep[][]{BOLTON07b,FINKELSTEIN12a,KUHLEN12,ROBERTSON15,BOUWENS15a,BOUWENS15,MITRA15,KHAIRE16,PRICE16}.	
	Simulations do not agree on $\fesc$ at high redshifts and find either very high \citep[e.g.,][]{SHARMA16} or very low values \citep[e.g.,][]{GNEDIN08,MA15}. Furthermore, they predict a strong dependence on dark matter halo mass and star formation \citep[e.g.,][]{WISE09,RAZOUMOV10}.
	 Direct observational constraints on $\fesc$ in the EoR are not possible because of the increasing opacity of the IGM to LyC photons at $z>4$ \citep[e.g.,][]{MADAU95,INOUE14}.
	Except one strong LyC emitter at $z=3.2$ with $\fesc>50\%$ \citep[][]{BARROS16,VANZELLA16}, the handful of confirmed LyC emitters at $z<3$ show all consistently $\fesc \lesssim 8\%$ \citep[][]{STEIDEL01,LEITET13,BORTHAKUR14,COOKE14,SIANA15,IZOTOV16a,IZOTOV16b,SMITH16,LEITHERER16}.
	The numerous non-detections listed in the literature show upper limits of $\fesc\sim2\%-5\%$ over large redshift ranges \citep[][]{VANZELLA10,SANDBERG15,GRAZIAN16,RUTKOWSKI16,GUAITA16,VASEI16}.
	If galaxies are responsible for ionizing the universe at $z>6$, clearly, their population averaged LyC escape fraction needs to increase substantially with redshift by at least a factor of two \citep[see also][]{INOUE06}.
	\textit{What methods can we use to access $\fesc$ observationally in the EoR?}
	Radiative transfer models suggest a correlation between the ratio of \oiii/\oii~and $\fesc$ in density bound \hii~regions \citep[e.g.,][]{NAKAJIMA14} and a handful of recent observational studies verify this positive correlation \citep[][]{BARROS16,VANZELLA16b,VANZELLA16,IZOTOV16a,IZOTOV16b}. The increased \oiii/\hbeta~ratios found in $z>5$ galaxies \citep[e.g.,][]{STANWAY14,ROBERTSBORSANI15,FAISST16a} hint towards an increasing \oiii/\oii~ratio for the global population of galaxies at high redshifts and therefore could be the smoking gun for a strong evolution in $\fesc(z)$.
	Currently, the \oii~line cannot be measured spectroscopically at $z>4$ and the use of broad-band photometry to determine \oii~line strengths is degenerate with the $4000\,{\rm \AA}$ Balmer break, a strong function of age and other galaxy parameters. However, local analogs of high redshift galaxies can be used to probe the physical properties of these galaxies.
	
	This paper aims to provide the first observationally based prediction of $\fesc$ in galaxies at $z>6$. To this end, we select local high-z analogs (LHAs) by their \halpha~emission \citep[see][]{FAISST16a}. We use these to predict the \oiii/\oii~ratios of high redshift galaxies, and, with an empirical correlation between \oiii/\oii~and $\fesc$, ultimately the redshift evolution of $\fesc$ (Section~\ref{seq:fesc}). With our prediction of $\fesc$ we then derive $\qhii(z)$ and $\tauel(z)$ and comment on the capability of galaxies to reionize the early universe (Section~\ref{seq:discussion}).
	Throughout this work we adopt a flat cosmology with $\Omega_{\Lambda,0}~=~0.7$, $\Omega_{m,0}~=~0.3$, and $h~=~0.7$. All stellar masses are scaled to a \citet{CHABRIER03} initial mass function.

	%%%%%%%%% FIGURE: LHA %%%%%%%%%
\begin{figure*}
\centering
\includegraphics[width=2.1\columnwidth, angle=0]{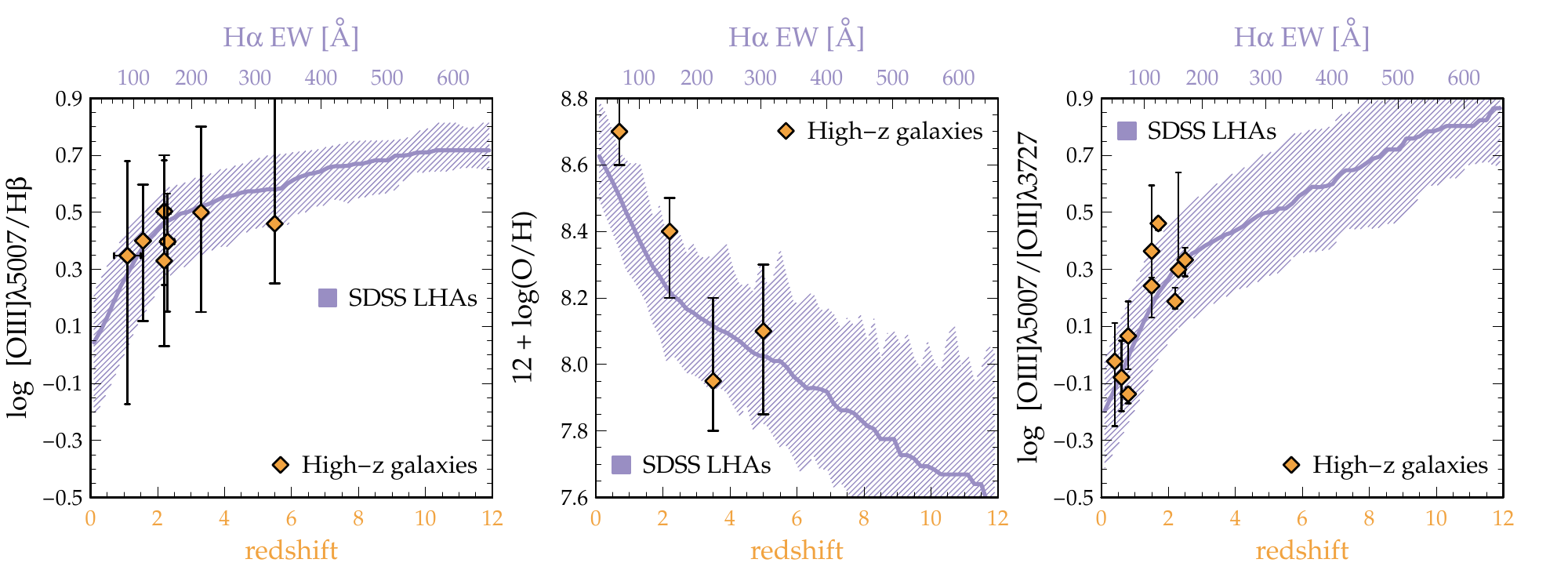}\\
\includegraphics[width=2.1\columnwidth, angle=0]{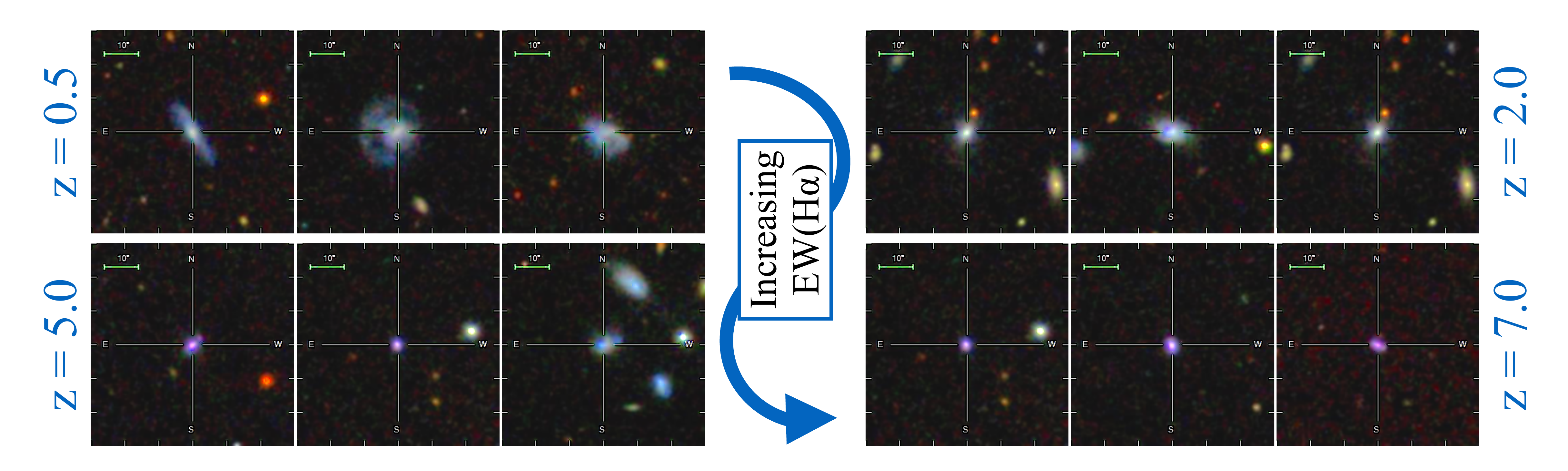}
%\vspace{4cm}
\caption{\textbf{Top:} The predicted redshift evolution of \oiii/\hbeta~(left), metallicity (middle), and \oiii/\oii~(right) from \halpha~EW selected LHAs in SDSS. The purple hatched band shows these properties as a function of $\ewha$ (upper $x$-axis) for the local SDSS galaxies. The orange symbols show these properties for intermediate and high redshift galaxies (see text for references) as a function of redshift (bottom $x$-axis). This comparison motivates the use of $\ewha$ selected local galaxies as analogs for high redshift galaxies. The connection between $\ewha$ and redshift is as given in Equation~\ref{eq:ewha} \citep[see also][]{FAISST16a}.
\textbf{Bottom:} SDSS color composites of three randomly picked representatives of LHAs for four different \halpha~EWs mimicking four different redshifts ($z=0.5$, 2.0, 5.0, 7.0 from left to right, top to bottom, stamps are $60\arcsec \times 60\arcsec$ in size). The LHAs show agreement in morphology with observed high redshift galaxies.
\label{fig:LHA}}
\end{figure*}
%%%%%%%%%%%%%%%%%%%%%%%%%%%%%%%%%%%%%%

\section{Predicting $\fesc$ at high redshifts}\label{seq:fesc}

%\section{Evolution of emission line properties from local high-z analogs}
\subsection{Locals as analogs for high redshift galaxies}\label{sec:analogs}

The resemblance of the physical properties of high redshift galaxies and sub-sets of galaxies at low ($z<1$) or local ($z\sim0$) redshifts has been known since almost a decade and it is subject of study in the very recent literature \citep[][]{CARDAMONE09,STANWAY14,GREIS16,BIAN16,FAISST16a,MASTERS16,ERB16}.
	Some of the most famous representatives of low redshift high-z analogs are the ``Green Peas'' at $z\sim0.2$ \citep[][]{CARDAMONE09} or the ultra strong emission line galaxies at $z\sim0.8$ \citep[USELs,][]{HU09}. In any case, the LHAs are characterized by an increased star-formation rate (SFR) surface density and \halpha~equivalent-width (EW)\footnote{Note that $\ewha$ is proportional to the specific SFR of a galaxy.} compared to the average local galaxy population \citep[e.g.,][]{MASTERS16}.
	In particular, \citet{FAISST16a} measure the \oiii/\hbeta~line ratios of average $z\sim5.5$ galaxies via the Spitzer color excess and verify a good agreement with LHAs selected by $\ewha > 300\,{\rm \AA}$. This first-order verification motivates the use of  \halpha~EW selected LHAs to predict spectroscopic properties of high redshift galaxies in the EoR.
	Here, we use a sample of more than $100,000$ local ($z<0.1$) galaxies drawn from the Sloan Digital Sky Survey \citep[SDSS,][]{YORK00} DR12 release \citep{ALAM15} using the SDSS query tool\footnote{\url{http://skyserver.sdss.org/dr12/en/tools/search/sql.aspx}}. The galaxies are selected to have ${\rm S/N} > 5$ in all the important optical emission lines (\oii, \oiii, \halpha, \hbeta, and \nii), and no AGN component.
%	, and stellar masses within $8.5 < \logm< 9.5$. The stellar mass range is designed to match the one of observed galaxies at high redshifts.
	We select LHAs for galaxies at a redshift $z$ by selecting SDSS galaxies with $\ewha|_{{\rm SDSS}} = \ewha(z) \pm \Delta(z)$, using the relation $\ewha(z)$ presented in \citet{FAISST16a} including the $1\sigma$ confidence interval ($\Delta$) in EW at a given $z$.
	The relation of $\ewha$ and redshift has been measured using various spectroscopic surveys at $z\sim0-3$ \citep[e.g.,][]{ERB06,STEIDEL14,SOBRAL15,SILVERMAN15} as well as up to $z\sim6$ \citep[e.g.,][]{FAISST16a} using the excess in Spitzer $[3.6\,{\rm \mu m}]-[4.5\,{\rm \mu m}]$ colors for a large sample of galaxies with spectroscopic redshift determinations as part of the \textit{Cosmic Evolution Survey} \citep[COSMOS,][]{SCOVILLE07}. For further details of the derivation of this relation, we refer the reader to \citet{FAISST16a}. Here we give the parameterization of this relation as
	
	\begin{equation}\label{eq:ewha}
	{\rm EW}({\rm H\alpha})(z) = 
	\begin{cases}
	20^{+15}_{-8} \times (1+z)^{1.87}, & z<2.2 \\
	37^{+51}_{-7} \times (1+z)^{1.30}, & z\geq2.2
	\end{cases}
	\end{equation}
	 
	 The error is given in the normalization and accounts for the physical scatter as well as the uncertainties of the measurements at low and high redshift. These uncertainties are propagated through this analysis and are included in the following results.
	We stress that, despite the obvious similarities of LHAs and intermediate galaxies ($z\sim2$, see also Figure~\ref{fig:LHA}), the use of LHAs to infer the properties of very high redshift galaxies has not been fully verified, yet. The following results therefore strongly depend on the assumption that strong \halpha~emitting local galaxies (equivalent to high specific SFR) are indeed similar to actual high-z galaxies and that the ISM properties do not greatly depend on the environment a galaxy was formed in. This does not have to be the case, since the cradles of formation for very high-z galaxies surely are different (more dense, more galaxy interactions) compared to the ones of local galaxies. Ultimately, the \textit{James Webb Space Telescope} (JWST) will be able to test these assumptions further and will provide a more clear picture.

	\subsection{Predicted emission line ratios of high-z galaxies}

	The top panels of Figure~\ref{fig:LHA} show the dependence of the spectroscopic properties of the local SDSS galaxies on $\ewha$ for galaxies with $8.5 < \logm < 9.5$ with a median of $\logm \sim 9.0$ (similar to the galaxies observed at high redshift). The stellar mass dependence of these relations is discussed in Section~\ref{sec:fescz}.
	We show the dependence of \oiii/\hbeta~ratio (left), gas phase metallicity\footnote{Metallicities are shown in the \citet{MAIOLINO08} calibration.} (middle), and \oiii/\oii~ratio (right) on $\ewha$ (top $x$-axis), with the median shown as purple line and the $1\sigma$ scatter visualized by the purple hatched band. Together with Equation~\ref{eq:ewha} and the assumption that these galaxies are high-z galaxy analogs, this can be interpreted as a redshift evolution of these quantities (bottom $x$-axis), which allows us to predict the spectroscopic properties of galaxies at higher redshifts. The open orange symbols show the measurement of the three quantities for actual galaxies at high redshift (from either spectroscopy or Spitzer color excess at $z>4$) from the literature (for \oiii/\hbeta: \citet{COLBERT13,STEIDEL14,SANDERS16,SILVERMAN15}; for metallicity: \citet{MAIOLINO08,FAISST16b}; for \oiii/\oii: \citet{RIGBY11,LEFEVRE13,DELOSREYES15,HAYASHI15,KHOSTOVAN16}).
	All in all, there is a good agreement in all the shown spectral properties of LHAs purely selected by $\ewha$ and actual high redshift galaxies up to $z\sim5$, where current measurements of \oii, \oiii~and \hbeta~are possible. This suggests that the \halpha~EW (closely related to the specific SFR) is strongly correlated with the conditions of the ISM in these galaxies, or, vice versa, the ISM of galaxies with strong \halpha~emission is very similar at all redshifts at least up to $z\sim5$. 
	Under the assumption of $\ewha$~being the main diagnostics of the spectral properties of galaxies, we use it as the quantity for the selection of local galaxies to predict the spectral properties of galaxies at $z>5$ where currently no such measurements are possible. As mentioned in Section~\ref{sec:analogs}, this assumption has to be tested, yet, by the next generation of telescopes such as \textit{JWST}.
	
	From the LHAs we infer \textit{average} \oiii/\hbeta~ratios of $\sim 4-5$ and \oiii/\oii~ratios larger than $3-4$ at $6 < z < 8$. The gas-phase metallicities of $z>6$ galaxies are predicted to be $12+\log(O/H) < 8.0$ on average, but with a substantial scatter leading to values of above 8.0 for some galaxies. Such large scatter is consistent with measurements of metallicity in $z\sim5$ galaxies based on rest-UV absorption features and is expected from the different evolutionary stages and dust attenuation as well as gas inflows in these systems \citep[e.g.,][]{FAISST16b}.
	
	Finally, the bottom panels of Figure~\ref{fig:LHA} show the morphological resemblance of our LHAs to high redshift galaxies. With increasing $\ewha$ (and therefore corresponding redshift), the LHAs become more compact and blue and show clumps in UV light as it is seen in high redshift galaxies at $z=2-4$ \citep[e.g.,][]{FORSTERSCHREIBER11,HEMMATI15}.

	%%%%%%%%% FIGURE: fesc vs. OIII/OII %%%%%%%%%
\begin{figure}
\centering
\includegraphics[width=1.0\columnwidth, angle=0]{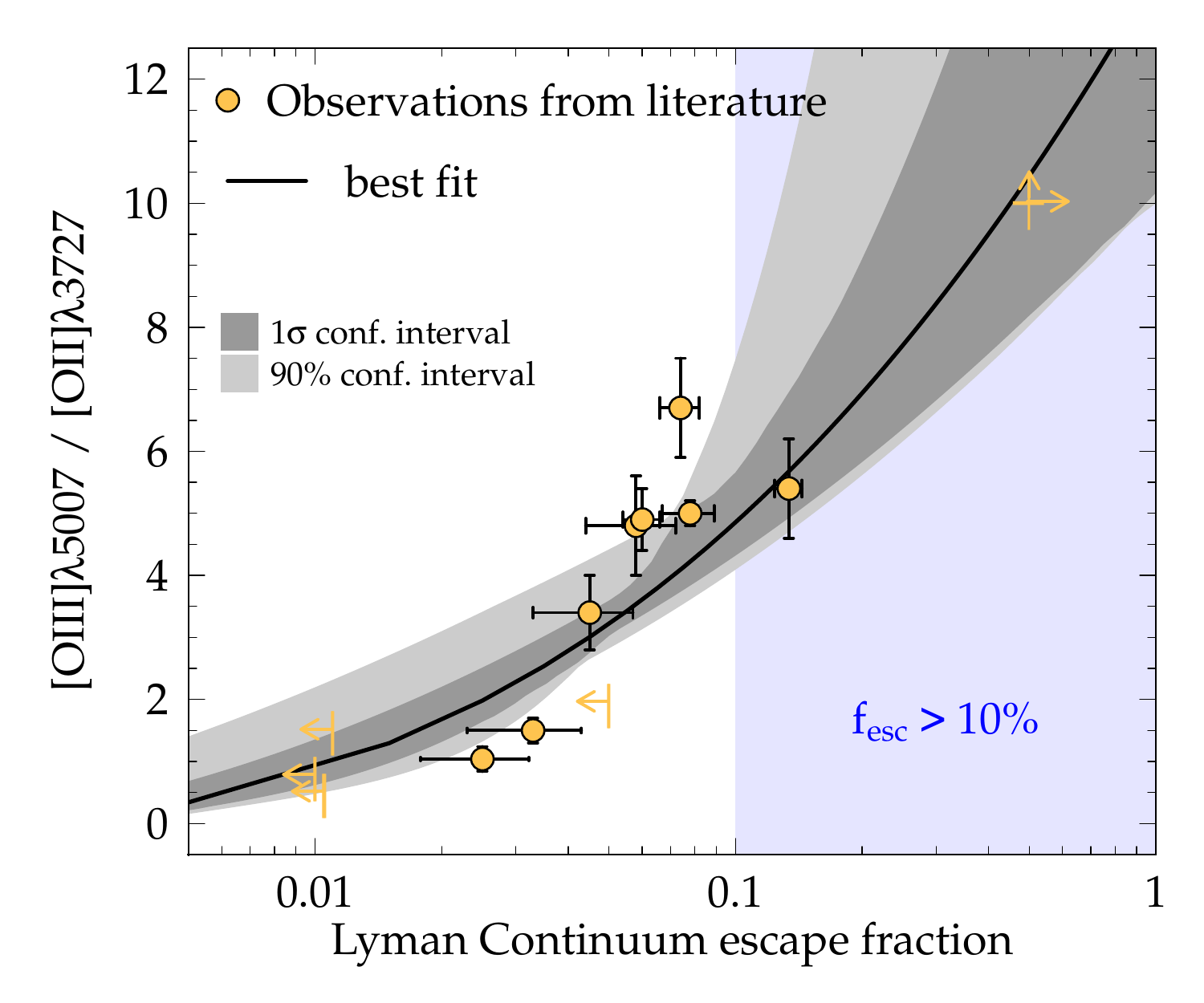}
%\vspace{4cm}
\caption{Observed correlation between \oiii/\oii~ratio with $\fesc$ from a compilation of literature data with limits shown by arrows. The black line shows the best fit median relation with $1\sigma$ and $90\%$ confidence interval envelope. The analytical parametrization is given in Equation~\ref{eq:fit}. The blue shaded area shows $\fesc>10\%$, needed for galaxies at $z>6$ to reionize the universe.
\label{fig:fescoiiioii}}
\end{figure}
%%%%%%%%%%%%%%%%%%%%%%%%%%%%%%%%%%%%%%

%%%%%%%%% FIGURE: fesc(z) %%%%%%%%%
\begin{figure*}
\centering
\includegraphics[width=2.1\columnwidth, angle=0]{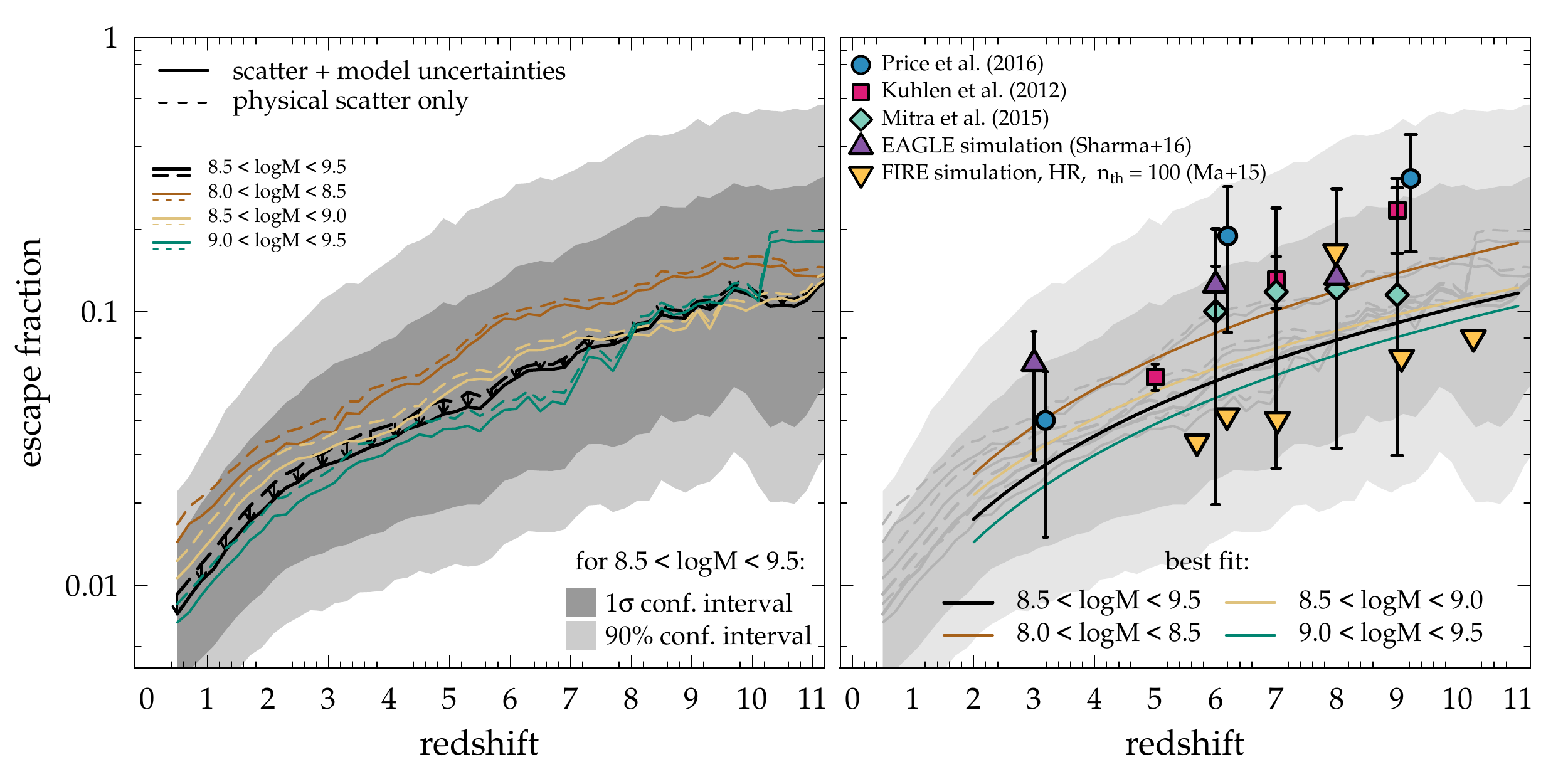}
%\vspace{4cm}
\caption{The redshift evolution of $\fesc$ infered by our LHAs.
% ($1\sigma$ and $90\%$ confidence intervals are shown as dark and light gray bands).
\textbf{Left:} The thick, black solid line shows the prediction for galaxies with $8.5 < \logm < 9.5$ together with $1\sigma$ (dark gray) and $90\%$ (light gray) confidence intervals including observed physical scatter in the \oiii/\oii~vs. redshift relation and uncertainties in the \oiii/\oii~vs. $\fesc$ fit. The dashed black line, shows the prediction \textit{only} including observational scatter (assuming no uncertainty in \oiii/\oii~vs. $\fesc$). The colored solid and dashed lines show the same for $8.0 < \logm < 8.5$ (brown), $8.5 < \logm < 9.0$ (beige), $9.0 < \logm < 9.5$ (green).
\textbf{Right:} Comparison of our predicted $\fesc(z)$ to model reconstructions \citep[][]{KUHLEN12,MITRA15,PRICE16} and simulations \citep[][]{SHARMA16,MA15} from the literature. The color lines show the best fits to our predictions parametrized by $\fesc(z>2) = f_{{\rm esc,0}}\left((1+z)/3\right)^\alpha$ for $2 < z < 8$ (see Table~\ref{tab:fits}).
\label{fig:fescz}}
\end{figure*}
%%%%%%%%%%%%%%%%%%%%%%%%%%%%%%%%%%%%%%

\subsection{Correlation between \oiii/\oii~and $\fesc$}\label{sec:fesc}

	The absorption of Lyman continuum photons in the IGM increases quickly by a factor of $100$ or more close to $z\sim4$ \citep[e.g.,][]{INOUE14}. The direct measurement of the galaxy intrinsic $\fesc$ at redshift greater than this is therefore not possible. However, its theoretically and observationally motivated connection with the \oiii/\oii~ratio may allow us to make predictions of $\fesc$ for distant galaxies.
	
	Commonly, $\fesc$ is measured in local galaxies from spectra or at intermediate redshifts by the detection of excess flux in narrow-band filters at rest-frame $\lambda < 900\,{\rm \AA}$. As summarized in Section~\ref{sec:intro}, the detection of Lyman continuum photons turns out to be difficult and current searches are mostly ending in non-detections. With the recent addition of Lyman continuum detections in mostly local galaxies, the positive correlation between $\fesc$ and the \oiii/\oii~ratio became observationally clear. The \oiii/\oii~ratio is a measure of the ionization parameter in galaxies, which correlates with the star-formation density and thus production of UV photons.
	A positive correlation is expected from radiative transfer simulations and is physically motivated by density-bound \hii~regions and stronger radiation fields prevailing in high redshift galaxies. In such environments, an increase in \oiii~flux, at a roughly constant \oii~emission, is expected in connection with a large amount of escaping ionizing photons and therefore high $\fesc$ \citep[e.g.,][]{NAKAJIMA14}.
	In Figure~\ref{fig:fescoiiioii}, we show 8 detections of $\fesc$ and 4 upper limits, each of them with reliable spectroscopic measurements of \oii~and \oiii~\citep[][]{LEITET13,BORTHAKUR14,IZOTOV16a,IZOTOV16b,VANZELLA16,LEITHERER16}. The limit at \oiii/\oii~$>10$ (estimated from only a spectroscopic detection of \oiii) and $\fesc>50\%$ shows the recent detection at $z=3.2$ \citep{BARROS16,VANZELLA16}.
	The positive correlation between $\fesc$ and \oiii/\oii~ratio is evident, although mostly driven by the limit at large $\fesc$. It should therefore be used with caution and its uncertainty must be included in the following analysis. We describe this relation with the analytical form
\begin{equation}
\label{eq:fit}
{\rm \oiii/\oii} = \phi + \left(\phi + a\right)^b + a,
\end{equation}
	and use the substitution $\phi = \log(f_{esc})$.
	The best fit ($a~=~2.60^{+0.09}_{-0.09}$ and $b~=~2.52^{+0.83}_{-0.16}$) is shown as dashed line. The gray band shows the $1\sigma$ and $90\%$ \textit{asymmetric} confidence interval of the fit, which is determined by a bootstrapping method and takes into account the limits and uncertainties of the measurements. We will see that including the uncertainties of this relation lowers the predicted $\fesc$ at a given redshift.
	The blue region in Figure~\ref{fig:fescoiiioii} shows $\fesc>10\%$, i.e., the value that must be reached by $z\sim6$ such that galaxies can dominate reionization under the current knowledge. We see that these values are reached at \oiii/\oii$\sim 4-7$, corresponding to $z\gtrsim 6.5$ for average galaxies at $\logm \sim 9.0$ (Figure~\ref{fig:LHA}).
	Finally, we note that the derivation of $\fesc$ itself depends on model assumptions. Specifically, the inclusion of binary stellar population, suggested to be more common at high redshifts \citep[][]{STANWAY14,STEIDEL16}, could lower the derived $\fesc$ values.

	\subsection{Redshift evolution and mass dependence of $\fesc$}\label{sec:fescz}
	
	Using the redshift evolution of \oiii/\oii~inferred from SDSS galaxies (left panel of Figure~\ref{fig:LHA}) and the empirical relation between \oiii/\oii~and $\fesc$ (Equation~\ref{eq:fit}, Figure~\ref{fig:fescoiiioii}), we can now predict $\fesc$ as a function of redshift under the given uncertainties. For this end, we use a Monte-Carlo sampling approach taking into account the scatter in the \oiii/\oii~vs. redshift relation and the uncertainties/limits in the \oiii/\oii~vs. $\fesc$ correlation.
	In detail, we sample 5000 galaxies for each redshift bin and draw \oiii/\oii~ratios to reproduce the observed distribution at a given redshift. For each of these galaxies we then draw $\fesc$ from the corresponding \oiii/\oii~ratio, which distribution we approximate with a skewed gaussian to take into account its asymmetric uncertainties.
	
	The left panel of Figure~\ref{fig:fescz} shows the final distribution of $\fesc(z)$ for different stellar mass bins. The dashed lines \textit{only} include the physical scatter in the \oiii/\oii~vs. redshift relation from our LHAs. The solid lines include the physical scatter \textit{and} the uncertainty in the modeling of the \oiii/\oii~vs. $\fesc$ relation. This in general \textit{lowers} the predicted $\fesc$ values (indicated by the arrows) and better constraints on the correlation between \oiii/\oii~and $\fesc$ are therefore crucial for a more detailed analysis. We also show the $1\sigma$ ($90\%$) confidence interval of the prediction as dark (light) gray band for the stellar mass range $8.5 < \logm < 9.5$ but omit it for the other masses for sake of clarity.
	Due to the $10-15\,\%$ higher \oiii/\oii~ratios for $\sim0.5\,{\rm dex}$ lower stellar masses \citep[see also][]{MASTERS16}, $\fesc$ shows a \textit{negative} correlation with stellar mass. In general, $\fesc$ is about $50\%$ higher per $0.5\,{\rm dex}$ smaller stellar mass in the range $8.0 < \logm < 9.5$ at $z\sim6$.
	From the predicted \oiii/\oii~ratios of local galaxies and the empirical relation between \oiii/\oii~and $\fesc$, we infer $\fesc=5.7_{-3.3}^{+8.3}\,\%$ at $z=6$ and $\fesc=10.4_{-6.3}^{+15.5}\,\%$ at $z=9$ for $8.5 < \logm < 9.5$ on average (errors given as $1\sigma$).
	Statistically, about $30\%$ of the galaxies at $z\sim6$ show $\fesc>10\%$, while this fraction becomes $50\%$ at $z\sim9$. We fit $\fesc = f_{{\rm esc,0}}\left(1+z\right)^\alpha$ for $2 < z < 8$ with $ f_{{\rm esc,0}}=2.3\pm0.1\,\%$ and $\alpha=1.17\pm0.02$ in the same stellar mass range. The best fits for the different stellar mass bins in the same redshift bin are listed in Table \ref{tab:fits}.
	The \textbf{right} panel of Figure~\ref{fig:fescz} compares our prediction of $\fesc(z)$ with measurements in the literature from simulations \citep[][]{MA15,SHARMA16} and $\fesc$ reconstructions from \textit{Planck}, Lyman-$\alpha$, and QSO data \citep[][]{KUHLEN12,MITRA15,PRICE16}. In general, our predicted $\fesc$ values are lower compared to other studies, except for the results from the \textit{FIRE} simulation \citep[][]{MA15}. Other simulations including supernova feedback (important in shaping the ISM and $\fesc$) suggest very similar results for $\fesc$ in the range of $10\%$ to $20\%$ at $z\sim9$ for our lowest stellar mass bin \citep[][]{KIMM14,CEN15}.
	All in all, our lowest stellar mass bin is consistent within $1\sigma$ with the literature.

	It has to be kept in mind that there is a strong dependence of $\fesc$ on stellar mass as described above. Simulations suggest a close correlation between dark matter halo mass or virial mass and the LyC escape fraction in relative agreement with the trends we find \citep[e.g.,][]{WISE09,RAZOUMOV10,KIMM14}. Furthermore, a negative correlation between \lya escape fraction and stellar mass is found \citep[][]{OYARZUN16}, which suggests a negative correlation between $\fesc$ and stellar mass via the close correlation between \lya and LyC escape fraction \citep[][]{DIJKSTRA16}.

%%%%%%% TABLE: FITS TO FESC redshift evolution %%%%
\begin{deluxetable}{ccc}
\tablecaption{Fit to predicted $\fesc(z)$ for different stellar mass bins.\label{tab:fits}}
\tablewidth{0pt}
\tablehead{
\colhead{stellar mass} & \colhead{$f_{{\rm esc,0}}$} $[\%]$ & \colhead{$\alpha$}
}
\startdata\\
$8.5 < \logm < 9.5$ & $1.7^{+0.1}_{-0.2}$ & $1.37^{+0.11}_{-0.10}$\\
$8.0 < \logm < 8.5$ & $2.5^{+0.2}_{-0.2}$ & $1.40^{+0.13}_{-0.10}$\\
$8.5 < \logm < 9.0$ & $2.1^{+0.2}_{-0.2}$ & $1.26^{+0.11}_{-0.10}$\\
$9.0 < \logm < 9.5$ & $1.4^{+0.4}_{-0.3}$ & $1.43^{+0.29}_{-0.41}$
\enddata
\tablenotetext{}{Analytical expression: $\fesc(z) = f_{{\rm esc,0}}\left((1+z)/3\right)^\alpha$. The fit is performed between $2 < z < 8$.}
\end{deluxetable}

%%%%%%%%%%%%%%%%%

\section{Can galaxies reionize the universe?}\label{seq:discussion}
	
	We use the LHAs to predict $\fesc(z)$ and find $\left<\fesc\right> \sim 6\%$ at $z=6$ and $\left<\fesc\right> \sim 10\%$ at $z=9$ for the stellar mass range $8.5 < \logm < 9.5$.
	This prediction comes with a substantial physical scatter (due to the scatter in the \oiii/\oii~ratios at a given redshift) and uncertainty stemming from the poorly constrained \oiii/\oii~vs. $\fesc$ relation. Statistically, $\sim30\%$ of the galaxies show $\fesc>10\%$ by $z=6$ for $8.5 < \logm < 9.5$, however, there is a stellar mass dependence that increases $\fesc$ by roughly $50\%$ per $0.5\,{\rm dex}$ decrease in stellar mass (see Figure~\ref{fig:fescz}).

	\textit{Is this enough for galaxies alone to reionize the universe?}
	The recent study by \citet[][]{PRICE16} reconstructs $\fesc(z)$ needed for reionization from the latest \textit{Planck} data and finds $2-3$ times higher $\fesc$ values at $z>6$ compared to our predictions. Their findings are just consistent (within $1\sigma$) with our observation based predictions for $\fesc$ for the smallest stellar mass bin ($8.0 < \logm < 8.5$). This suggests that the commonly found galaxies at high redshifts with $\logm \sim9.0$ are not sufficient to ionize the early universe, instead it is mostly driven by low-mass, low-luminosity galaxies at $\logm \sim 8.0$.
	
	In the following, we want to investigate the above question in more detail and derive two important quantities: $\qhii(z)$ (the volume fraction of ionized hydrogen) and $\tauel(z)$ (the integrated electron scattering optical depth). To achieve this end, several assumption have to be made.
	First, the faint end cut off of the UV LF ($\muvlim$) determines the number of faint galaxies that are available for ionization (similar to the stellar mass function, remember, $\fesc$ is anti-correlated with stellar mass). High redshift galaxies lensed by foreground low redshift galaxy clusters allow us to probe the UV LF to very faint magnitudes of $\muv\sim-12$ at $z = 6-8$ and have shown no indication of a turn-over \citep[][]{ALAVI14,LIVERMORE16}. It is therefore safe to assume a value between $-13 < \muvlim < -10$.
	Furthermore, the Lyman continuum photon production efficiency ($\xiion$) and the clumping factor ($C$) need to be known. The former is measured observationally and is found to be $\log(\xi_{{\rm ion}}/[{\rm Hz~erg^{-1}}]) = 25.4\pm0.1$ for a wide range of galaxy properties at $z\sim5$ \citep{BOUWENS15}. The clumping factor $C=\left< n_{H}^2\right> / \bar{n}_{H}^2$ is proportional to the recombination rate of hydrogen\footnote{The recombination rate is proportional to the hydrogen density squared.} and thus the net production rate of ions. It is commonly constrained from simulations to be between 2 and 5, and we assume $\left< C \right> = 3$ \citep[e.g.,][]{FINLATOR12}. Other reasonable values of $C$ have little impact on the following results.

%%%%%%%%% FIGURE: final, 2 panels: QHII(z) and tau(z) with Planck %%%%%%%%%
\begin{figure*}
\centering
\includegraphics[width=1.05\columnwidth, angle=0]{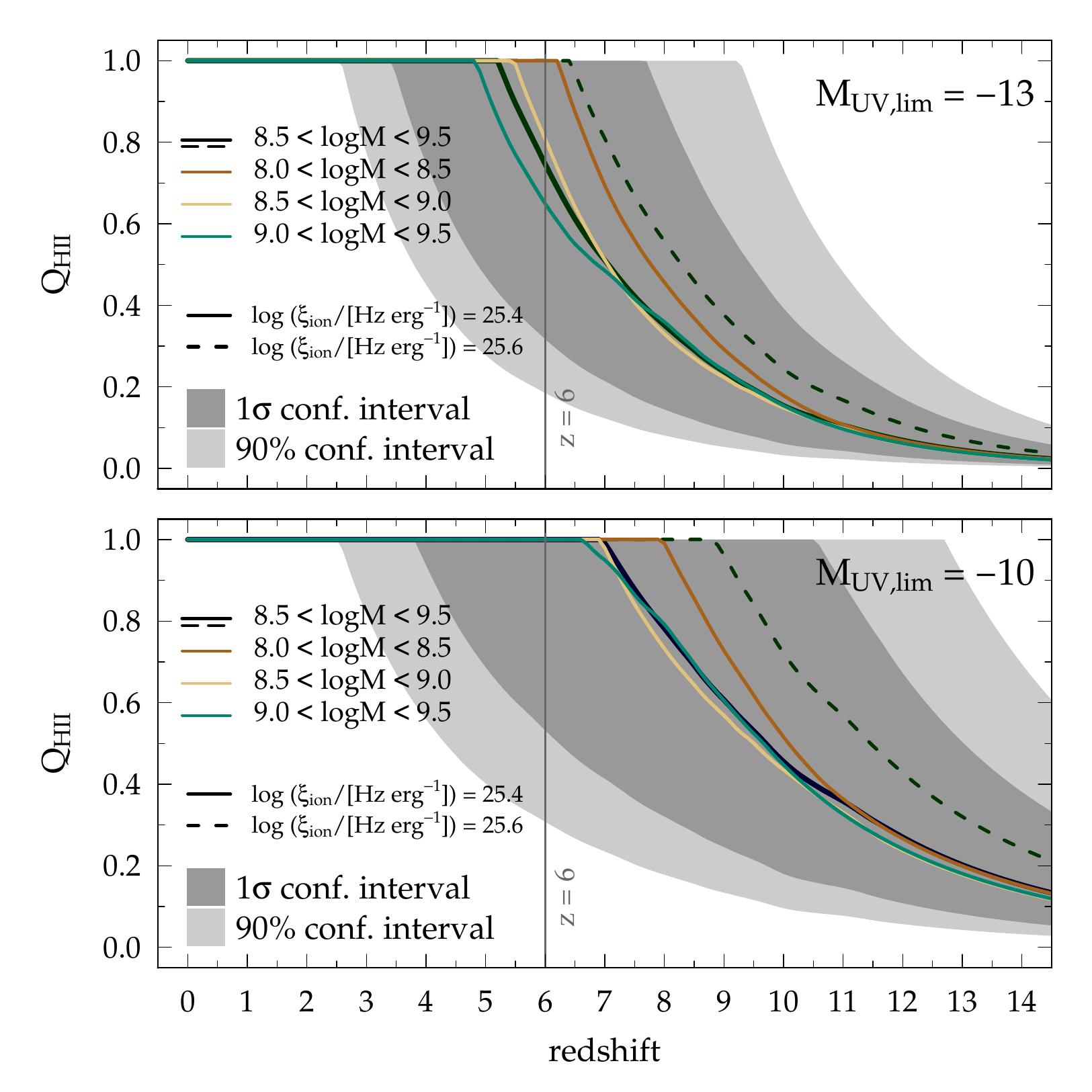}
\includegraphics[width=1.05\columnwidth, angle=0]{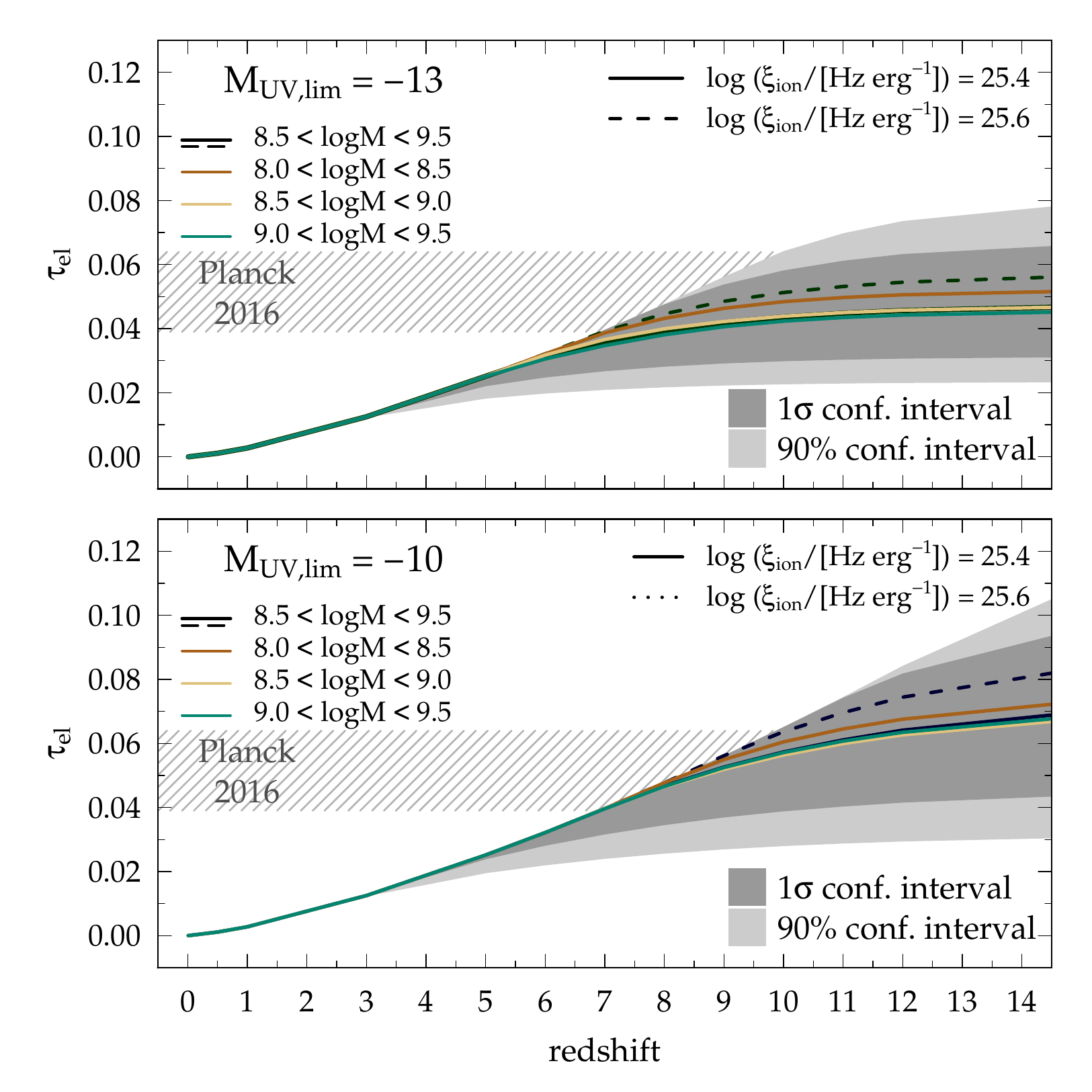}
%\vspace{4cm}
\caption{
\textbf{Left:} Volume fraction of ionized hydrogen as a function of redshift assuming $\log\left(\xi_{\rm ion} \right)=25.4$ for $\muvlim=-13$ (top) and $\muvlim=-10$ (bottom) with $1\sigma$ and $90\%$ confidence intervals (for $8.5 < \logm < 9.5$, only). The different stellar mass bins are indicated with colors. The thin line marks $z<6$ when the universe is observed to be fully ionized. We also show $\log\left(\xi_{\rm ion} \right)=25.6$ for $8.5 < \logm < 9.5$ as dashed line for reference.
\textbf{Right:} Electron scattering optical depth integrated up to different redshifts for $\muvlim=-13$ (top) and $\muvlim=-10$ (bottom) with $1\sigma$ and $90\%$ confidence intervals (for $8.5 < \logm < 9.5$, only). The different stellar mass bins are indicated with colors. The gray hatched band marks the recent constraints from \textit{Planck}.  We also show $\log\left(\xi_{\rm ion} \right)=25.6$ for $8.5 < \logm < 9.5$ as dashed line. All in all, we find that our observational based prediction of $\fesc$ are just enough for galaxies to reionize the universe by $z\sim5.3$. Lower mass galaxies are predicted to reionize the universe by $\Delta z \sim 0.5$ earlier.
\label{fig:final}}
\end{figure*}
%%%%%%%%%%%%%%%%%%%%%%%%%%%%%%%%%%%%%%	

	With the relatively good constraints on $\xiion$ and $C$ and our empirical prediction of $\fesc(z)$ we can now derive $\qhii(z)$ and $\tauel(z)$ using the following set of basic equations.
	
\begin{equation}
\label{eq:tauel}
\tau_{{\rm el}}(z) = c \left< n_{{\rm H}}\right> \sigma_{{\rm T}} \int^z_0 f_{{\rm e}} Q_{{\rm HII}}(z') H^{-1}(z') (1+z')^2 dz'
\end{equation}

\begin{equation}
\label{eq:qhii}
\dot{Q}_{{\rm HII}} = \frac{\dot{n}_{{\rm ion}} }{\left< n_{\rm H} \right>} - \frac{Q_{{\rm HII}}}{t_{{\rm rec}}}
\end{equation}

\begin{equation}
\label{eq:ndot}
\dot{n}_{{\rm ion}} = \fesc \xi_{{\rm ion}} \rho_{{\rm uv}}
\end{equation}

\begin{equation}
\label{eq:trec}
t_{\rm rec} = \left[ C\,\alpha_{B}(T) (1+Y_{p}/4{X_{p}}) \left< n_{{\rm H}} \right> (1+z)^{3}  \right]^{-1}
\end{equation}

\begin{equation}
\label{eq:ab}
\alpha_{B} = 2.6 \times 10^{-13}  \left(\frac{T}{10^4 {\rm K}}\right)^{-0.76} {\rm cm^3/s}
\end{equation}

\begin{equation}
\label{eq:nh}
\left< n_{H} \right> =  1.67 \times 10^{-7} \left(\frac{\Omega_{b}h^2}{0.02}\right) \left(\frac{X_{p}}{0.75}\right) {\rm cm^{-3}}
\end{equation}

where $t_{rec}$ is the hydrogen recombination time with $\alpha_{B}$ the case B recombination coefficient. We assume $X_{p} = 0.75$ for the hydrogen mass fraction \citep[e.g.,][]{HOU11}, the helium mass fraction is given as $Y_p = 1-X_p$ \citep[][]{KUHLEN12}, and a fraction of free electrons as $f_e = 1+Y_p/2X_p$ at $z\leq4$ and $f_e = 1+Y_p/4X_p$ at $z>4$. Furthermore, we use a baryon density $\Omega_{b}=0.04$, the Thompson scattering cross-section $\sigma_{\rm T}=6.653\times10^{-25}\,{\rm cm^2}$, and an IGM temperature $T=20,000\,{\rm K}$. We assume $C=3$ and $\log\left(\xi_{{\rm ion}}\right)=25.4$ and $25.6$. For the integrated UV luminosity density, $\rho_{\rm uv}$, we use $\muvlim=-13$ and $-10$ and the UV luminosity functions by \citet{MASON15}.

	With our observationally driven prediction of $\fesc(z)$ and our reasonable assumptions for $C$ and $\xiion$, we can now investigate whether galaxies can reionize the universe by $z\sim6$.
	The left panels of Figure~\ref{fig:final} show $\qhii(z)$ for $\muvlim\,=\,-13$ (top) and $-10$ (bottom). The $1\sigma$ and $90\%$ confidence intervals from our $\fesc$ predictions are given for $\log\left(\xiion\right)\,=\,25.4$ and $8.5 < \logm < 9.5$. We also show $\log\left(\xiion\right)\,=\,25.6$ as dashed line for reference. The population averaged ($8.5 < \logm < 9.5$) results are shown in black together with other stellar mass bins with colors as in Figure~\ref{fig:fescz}.
	The right panels of Figure~\ref{fig:final} show $\tauel(z)$ with the same color coding and assumptions.
	We find that galaxies with $\logm\sim9.0$ are capable of ionizing the IGM by $z_{\rm ion}=5.3_{+2.4}^{-1.8}$ and yield $\tauel\sim0.05$ with the combination $\left(\muvlim,\log\xiion \right)=\left(-13,25.4 \right)$. Note that this is a population averaged quantity and single galaxies may show very different escape fractions and therefore contribute to a non-isotropic reionization of the universe. Particularly, due to the negative correlation between \oiii/\oii~and stellar mass, a population of low mass galaxies ($\logm \sim 8.0-8.5$) will reionize the universe slightly earlier ($\Delta z \sim 0.5$, so roughly at $z\sim6$).
	These findings are in good agreement with measurement of \lya forest transmission, quasar absorption, and gamma-ray bursts \citep[e.g.,][]{FAN06,TOTANI06,BOLTON07a,MCQUINN08,FAUCHERGIGUERE08,CARILLI10,BOLTON11,MCGREER11,MORTLOCK11,SCHROEDER13} as well as the constraint from \textit{Planck} on the electron scattering optical depth (upper right panel of Figure~\ref{fig:final}).
	The combination $\left(\muvlim,\log\xiion \right)=\left(-10,25.4 \right)$ yields $z_{\rm ion}=6.9_{+3.5}^{-3.1}$, which is also in agreement within uncertainties with the complementary observational constraints, but it overshoots the constraints on $\tauel$ by \textit{Planck}; it leads to a too early reionization.
	This analysis also depends on the assumed value for $\xiion$ (see dashed line in Figure~\ref{fig:final} showing $\log\left(\xiion\right) = 25.6$) and in particular higher $\xiion$ have the same effect as lowering stellar mass and lead to an earlier reionization.
	Several recent studies have suggested that the inclusion of binary stars in the stellar population models may result in higher $\xiion$ than the current canonical value of $\log\left(\xiion\right) \sim 25.5$ \citep[][]{STANWAY16,WILKINS16}. In fact, binary models are expected to be more suitable at high redshifts because of the low metallicity environments and young stellar populations \citep[][]{MA16,STEIDEL16,WILKINS16}. If binary populations were prevalent at high redshifts, this would lower the necessary $\fesc$ for galaxies to keep to universe ionized at $z<9$ to modest values of $4-24\%$, comparable to our estimates \citep[e.g.,][]{STANWAY16}.

	 Finally, there are not many constraints on the dependency of $\xiion$ on other galaxy properties. However, \citet{BOUWENS15} see a weak \textit{negative} trend between the UV continuum slope ($\beta$) and $\xiion$ in their data at $3.8 < z < 5.0$. Assuming a \textit{positive} correlation between stellar mass and $\beta$ \citep[i.e., more massive galaxies having shallower slopes and likely more dust, e.g.,][]{FINKELSTEIN12a,BOUWENS14}, this would suggest a \textit{smaller} $\xiion$ for more massive galaxies. These trends thus suggest that less massive galaxies might be even more efficient in ionizing, or, vice versa, more massive galaxies less.

\section{Conclusions}
	
	We use empirical trends seen in local galaxies to predict emission line ratios and Lyman continuum escape fractions at high redshifts.
	For this end, we combine the positive correlation of the \oiii/\oii~line ratio and $\fesc$ based on low redshift galaxies with the predicted \oiii/\oii~ratios at high redshift from LHAs.
	We find increasing \oiii/\oii~line ratios with increasing redshifts reaching values of \oiii/\oii~$\sim3-4$ commonly by $z=6$, which translates into $\fesc>6\%$ on average at $z>6$ for galaxies with $\logm \sim 9.0$. Statistically, including uncertainties and scatter, roughly $30\%$ of galaxies at $z=6$ show $\fesc>10\%$ and this fraction increases to $50\%$ at $z=9$. This first observation based prediction of $\fesc$ suggests that its values for high redshift galaxies are substantially higher than the currently low $\fesc$ limits measured in low-z galaxies and thus hints towards a strong redshift evolution of $\fesc$.
	However, we also find a strong stellar mass dependency of $\fesc$, driven by the stellar mass dependency of the \oiii/\oii~ratio. In particular, a decrease by $0.5\,{\rm dex}$ in stellar mass results in an \textit{increase} in $\fesc$ of $50\,\%$. Note that this is in agreement with the dark matter halo dependency of $\fesc$ predicted by various simulations. If true, the stellar mass function and its evolution with redshift -- that is likely coupled with the rest-UV luminosity function -- is an important ingredient to access the importance of galaxies in the EoR.
	With our observation based prediction for the population averaged $\fesc(z)$ (thus fixing an important previously unknown variable) and reasonable assumptions for $C$ and $\xiion$ we find that galaxies at $\logm \sim 9.0$ release a sufficiently large number of ionizing photons to reionize the universe by $z_{\rm ion}=5.3_{+2.4}^{-1.8}$ for $\muvlim=-13$ and $z_{\rm ion}=6.9_{+3.5}^{-3.1}$ for $\muvlim=-10$. Galaxies at lower masses are able to reionize the universe by $\Delta z \sim 0.5$ earlier.
	
	This work should be understood as the beginning of a more detailed study of $\fesc$ during the EoR. Until now, it is still hampered by the large uncertainties in the \oiii/\oii~vs. $\fesc$ relation. Furthermore, the link between LHA and actual very high redshift galaxies needs to be explored in more detail. Future spectroscopic observations by the \textit{Hubble Space Telescope} will enhance the sample sizes of galaxies with LyC detection and will add to a better understanding of the link between \oiii/\oii~and $\fesc$. Furthermore, \textit{WFIRST} and \textit{JWST} will ultimately measure \oiii~and \oii~in high redshift galaxies and thus verify the link between local galaxies and the first galaxies formed.

\acknowledgements
The author would like to thank Dan Masters, Peter Capak, Janice Lee, and Kirsten Larson for valuable discussions which improved this manuscript. Furthermore, the author would like to thank the anonymous referee for the very useful feedback. AF acknowledges support from the Swiss National Science Foundation.




\bibliographystyle{aasjournal}
\bibliography{bibli.bib}

%\begin{thebibliography}{}

%\end{thebibliography}



\end{document}